\documentclass[aps, prl, superscriptaddress, preprint]{revtex4-1}

\usepackage{amsmath}
\usepackage{amsfonts}
\usepackage{amssymb}
\usepackage{graphicx}
\usepackage{color}
\usepackage{enumerate}
\usepackage[final]{pdfpages}
\usepackage{lastpage}

\makeatletter
\AtBeginDocument{\let\LS@rot\@undefined}
\makeatother

\begin{document}

\title{Nucleation and growth bottleneck in the conductivity recovery dynamics of nickelate thin films}

\author{E. Abreu}
\email{elsabreu@phys.ethz.ch}
\affiliation{Institute for Quantum Electronics, Department of Physics, ETH Zurich, 8093 Zurich, Switzerland}
\author{D. Meyers}
\affiliation{Department of Materials Science and Engineering, University of California, Berkeley, CA 94720, USA}
\author{V. K. Thorsm\o lle}
\affiliation{Department of Physics, Boston University, Boston, MA 02215, USA}
\affiliation{Department of Physics, UC San Diego, La Jolla, CA 92093, USA}
\author{J. Zhang}
\affiliation{Department of Physics, UC San Diego, La Jolla, CA 92093, USA}
\affiliation{Department of Physics, The Hong Kong University of Science and Technology, Clear Water Bay, Kowloon, Hong Kong, China}
\author{X. Liu}
\affiliation{Department of Physics and Astronomy, Rutgers University, Piscataway, NJ 08854, USA}
\author{K. Geng}
\affiliation{Department of Physics, Boston University, Boston, MA 02215, USA}
\author{J. Chakhalian}
\affiliation{Department of Physics and Astronomy, Rutgers University, Piscataway, NJ 08854, USA}
\author{R. D. Averitt}
\affiliation{Department of Physics, UC San Diego, La Jolla, CA 92093, USA}

\begin{abstract}
We investigate THz conductivity dynamics in NdNiO$_3$ and EuNiO$_3$ thin films following a photoinduced thermal quench into the metallic state and reveal a clear contrast between first- and second-order dynamics.
While in EuNiO$_3$ the conductivity recovers exponentially, in NdNiO$_3$ the recovery is non-exponential and slower than a simple thermal model.
Crucially, it is consistent with first-order dynamics and well-described by a 2d Avrami model, with supercooling leading to metastable phase coexistence.
The large transients seen in our films are promising for fast electronic (and magnetic) switching applications.
\end{abstract}

\maketitle


Order parameter coupling is one of the hallmarks of transition metal oxides (TMOs), in which macroscopic properties arise from competition between microscopic interactions.
The richness of the resultant phase diagrams \cite{Dagotto2005} imbues these materials with extreme sensitivity to external perturbations which can be investigated using various time-resolved measurements \cite{Zhang2014}.
In particular, materials that exhibit insulator-to-metal transitions (IMTs) present an ongoing challenge, regarding both fundamental properties and photoinduced phase transitions \cite{IMT_dobro}.
Photoinduced IMTs have been investigated in a host of TMOs, with vanadates (VO$_2$ and V$_2$O$_3$) as canonical examples \cite{Cavalleri2001, Wegkamp2014, Morrison2014, Gray2018, Abreu2015, Singer2018}.
However, in order to better understand the fundamental aspects of photoinduced dynamics and the technological potential of IMTs in TMOs it is imperative to investigate materials with coupled charge, spin and structure orders.  

Rare earth nickelates, \emph{R}NiO$_3$, are excellent candidates for this purpose.
They exhibit a variety of electronic, magnetic and structural phases, accessible through changes in temperature and pressure, or by a variation of the tolerance factor of their characteristic distorted perovskite structure through a change of \emph{R} \cite{Torrance1992}.
In particular, \emph{R}NiO$_3$ with smaller \emph{R} radius, such as EuNiO$_3$ (ENO), exhibit two distinct second-order transitions associated with magnetic ordering (at the N\'eel temperature, $T_N$) and an IMT (at $T_{IMT}$), whereas large \emph{R} nickelates, such as NdNiO$_3$ (NNO), undergo a single first-order transition between an antiferromagnetic insulator and a paramagnetic metal.
This strongly suggests that a coupling exists between electronic and magnetic properties.
The coupling is expected to be \emph{R} dependent but its precise nature remains unclear, with an intriguing example being the recent observation of antiferromagnetic order in bulk metallic LaNiO$_3$ \cite{Guo2018}.
The rich phase diagram of \emph{R}NiO$_3$ has remained somewhat unexplored due to challenges in bulk crystal growth \cite{Torrance1992}.
Thin epitaxial films and heterostructures arose as a convenient alternative, enabling the investigation of intrinsic nickelate properties while also opening up an entire new functional landscape through tuning of e.g. epitaxial strain and interfacial effects \cite{Catalan2008, Boris2011, Liu2013, Middey2016}.

\begin{figure*} [htb]
\begin{center}
\includegraphics[width=1\textwidth,keepaspectratio=true]{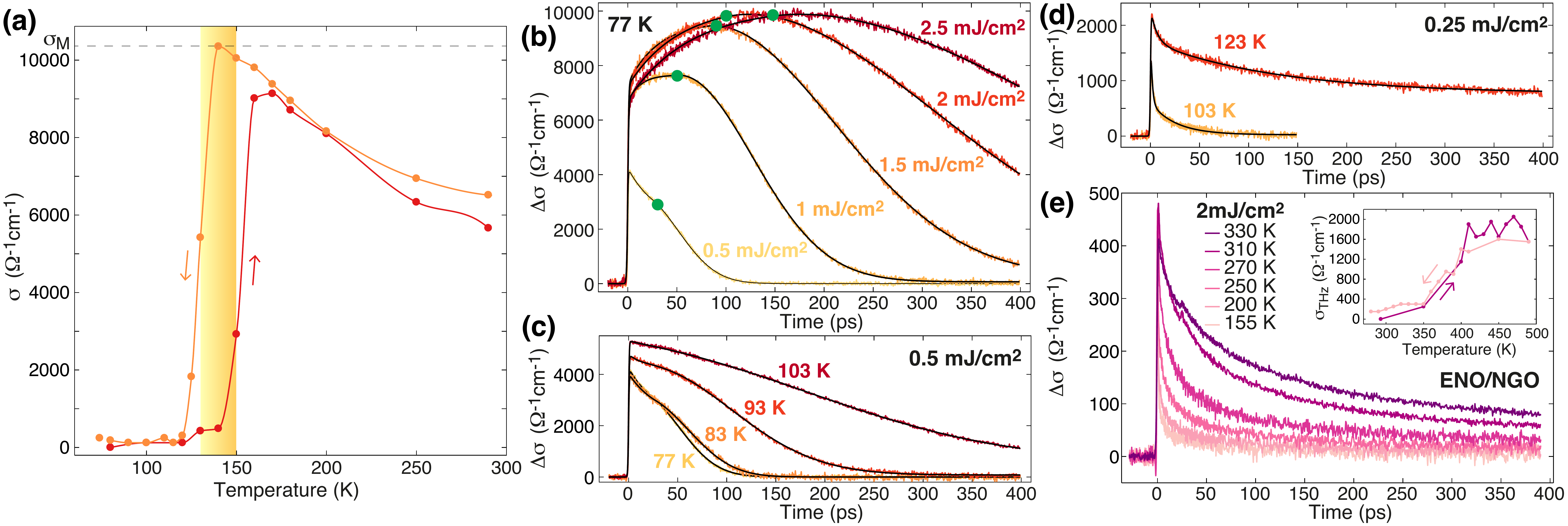}
\caption{
(a) Static THz conductivity in NNO, for increasing (red) and decreasing (orange) temperature.
Shading marks the coexistence of metallic and insulating domains.
(b, c, d) Transient THz conductivity in NNO vs. pump - probe delay time, for (b) fixed $T_i$, varying $F_{inc}$, and (c, d) fixed $F_{inc}$, varying $T_i$.
Thin black lines are fits to Eq. \ref{eq1}.
Green dots mark the onset ($t_d$) of the non-exponential recovery \cite{supMaterial}.
e) In ENO, $\Delta\sigma (t)$ for $F_{inc}=$ 2 mJ/cm$^2$, varying $T_i$.
Inset: $\sigma (T)$, yielding $T_{IMT} \simeq 380$ K.
}
\label{figOPTP}
\end{center}
\end{figure*}

In this work we investigate thin epitaxial films of NNO and ENO using optical pump - THz probe (OPTP) spectroscopy.
In the NNO films $T_{IMT} \simeq 140$ K (compared to 200 K in bulk samples due to strain in the films) \cite{Torrance1992, Liu2013}, while for the ENO films $T_N = 200$ K (as in bulk) and $T_{IMT} \simeq 380$ K (100 K lower than in bulk) \cite{Meyers2013, Torrance1992}.
Our investigation of NNO and ENO enables a direct comparison of first- and second-order IMT conductivity dynamics.
We note that previous time-resolved studies of the IMT in \emph{R}NiO$_3$ thin films focused on vibrational excitation of substrate phonon modes \cite{Caviglia2012, Forst2015, Forst2017} or were performed indirectly using an optical probe \cite{Ruello2007, Esposito2018}.
Magnetic \cite{Caviglia2013, Forst2015} and charge order dynamics \cite{Esposito2018} in photoexcited NNO films were also investigated, using time-resolved resonant soft x-ray diffraction.
Our conductivity dynamics are initiated with above bandgap photoexcitation which induces a thermal quench from the insulating to the metallic state.
Photoexcitation of NNO drives a prompt collapse ($\sim 1$ ps) into the metallic state followed by an additional conductivity increase upon cooling within the metallic phase.
Crucially, further cooling to below $T_{IMT}$ causes the conductivity to decrease exhibiting slowed down first-order dynamics, with supercooling leading to metastable phase coexistence and percolative recovery of the antiferromagnetic insulating state.
ENO also exhibits a prompt increase of the conductivity but the recovery follows the simple exponential relaxation expected for a continuous IMT, in stark contrast to the novel first-order dynamics observed in NNO.
Comparison of our results to the magnetic dynamics from Caviglia \emph{et al.} \cite{Caviglia2013} suggests that the collapse of magnetic order tracks the conductivity, and that a strong coupling exists between the magnetic and electronic order.


Thin films of NNO and ENO (15 u.c., $\sim 5.7$ nm)  were fabricated using layer-by-layer growth on [001]-oriented NdGaO$_3$, leading to 1.4 $\%$ and 1.5 $\%$ tensile strain, respectively \cite{Liu2013, Meyers2013}.
For the transient conductivity measurements we used $\sim$ 50 fs 1.55 eV pulses from a 3 mJ Ti:Sapphire amplified laser operating at 1 kHz.
THz probe pulses peaked at $\sim 1$ THz were generated by optical rectification and detected by electro-optic sampling in 1 mm thick ZnTe crystals.
The pump and probe beams impinged on the sample at approximately normal incidence, with a pump focal spot size of $\sim 5$ mm diameter FWHM, significantly larger than the probe.
The penetration depth of the pump and probe beams are both much larger than the film thickness.


We begin with our results on NNO.
First, we performed temperature dependent THz time domain spectroscopy measurements of the NNO film without photoexcitation.
The THz conductivity, $\sigma (T)$, is shown in Fig. \ref{figOPTP}(a), which effectively corresponds to the d.c. conductivity since the carrier scattering frequency lies beyond the measured spectral range (0.1 - 2.5 THz).
The IMT occurs at $T_{IMT} \simeq 140$ K, accompanied by hysteresis.
This is consistent with the first-order nature of the transition and reveals a $\sim 20$ K range of phase coexistence (shaded region in Fig. \ref{figOPTP}(a)).
The maximum conductivity, obtained around $T_{IMT}$, is $\sigma_M \simeq 10^4$ ($\Omega$ cm)$^{-1}$, comparable to bulk values \cite{Torrance1992}.
For higher temperatures the conductivity decreases significantly with increasing temperature.

Next, we use OPTP spectroscopy to measure the conductivity changes ($\Delta\sigma (t)$) arising from dynamic spectral weight transfer following excitation at 1.55 eV.
$\Delta \sigma (t)$ is plotted in Figs. \ref{figOPTP}(b) -  \ref{figOPTP}(d) for initial temperatures $T_i<T_{IMT}$ of 77 - 123 K, and incident pump fluences $F_{inc} = $ 0.25 - 2.5 mJ/cm$^2$.
For all fluences there is a fast initial rise of the conductivity, $\sim $ 1 ps, that occurs on the timescale of electron phonon (e-ph) thermalization.
The $\Delta \sigma (t)$ evolution after this photoinduced thermal quench depends strongly on $F_{inc}$ and $T_i$, which determine the proximity of the final temperature $T_f$  (immediately after e-ph thermalization) to $T_{IMT}$.
Figure \ref{figOPTP}(b) shows the dynamics for $T_i =$ 77 K over a range of fluences. 
For $F_{inc}=$ 0.5 mJ/cm$^2$, $\Delta\sigma (t)$ is $\sim 4000$ ($\Omega$ cm)$^{-1}$ at 1 ps, with a subsequent decrease and a slight shoulder at $\sim 30$ ps (marked with a green dot).
This last feature becomes much clearer for $F_{inc}=$ 1 - 2.5 mJ/cm$^2$.
For these higher fluences the thermal quench ($T_f$) is well above $T_{IMT}$, with the initial $\Delta\sigma (t)$ increase at 1 ps reaching $\sim 7000 - 8000$ ($\Omega$ cm)$^{-1}$.
A subsequent second rise in the conductivity peaks at later times with increasing fluence, as indicated with the green dot on each trace \cite{supMaterial}.
Finally, there is a non-exponential recovery, with a decay time that increases with fluence.
This deviation from a simple exponential recovery suggests the onset of first-order dynamics, as discussed in detail below.
With decreasing fluence the dynamics evolve towards an exponential recovery, as shown in Figs. \ref{figOPTP}(c) - \ref{figOPTP}(d).
Indeed, at $0.25$ mJ/cm$^2$ (Fig. \ref{figOPTP}(d)) there is a two-exponent decay of $\Delta \sigma (t)$, with a fast component, $\sim 5$ ps, and a longer decay time that increases as $T_i$ approaches $T_{IMT}$. 


Further insight into the $\Delta \sigma (t)$ dynamics in NNO can be obtained by fitting the data in Figs. \ref{figOPTP}(b) -  \ref{figOPTP}(d) to:
\begin{equation}
\label{eq1}
\begin{aligned}
\Delta \sigma_{fit} (t) = 
\frac{1}{e^{-\frac{t}{\tau_i}}+1} \big[ \Delta\sigma_i + \Delta\sigma_1 (e^{-\frac{t}{\tau_1}}-1) - \Delta\sigma_2 (e^{-\frac{t}{\tau_2}}-1) \big]
\big[d \; e^{-\big(\frac{t-t_d}{\tau_{Av}}\big)^2} + (1-d)\big],
\end{aligned}
\end{equation}

\noindent where $\tau_i$ and $\Delta\sigma_i$ correspond to the initial ultrafast conductivity increase.
At low $T_i$ and $F_{inc}$ (Fig. \ref{figOPTP}(d)), $\tau_1$ and $\Delta\sigma_1$ are the decay time and amplitude of the fast conductivity decrease while $\tau_2$ and $\Delta\sigma_2$ ($<$0) correspond to the slow decrease.
For large $T_i$ and $F_{inc}$ (Fig. \ref{figOPTP}(b)), $\tau_2$ and $\Delta\sigma_2$ ($>$0) are the rise time and amplitude of the second slow conductivity increase.
Note that for $F_{inc}=$ 0.5 mJ/cm$^2$ (Fig. \ref{figOPTP}(c)) there is still a small $\Delta\sigma_1$ recovery component, absent for higher fluences.
The term  in $\tau_{Av}$ corresponds to the non-exponential conductivity recovery, which has an associated delay, $t_d$, discussed below.
Finally, $d$ accounts for a small and long lived ($>1$ ns) decay component arising from the final stage of heat transfer to the NdGaO$_3$ substrate.

Fits to Eq. \ref{eq1} are superimposed onto the data in Figs. \ref{figOPTP}(b) - \ref{figOPTP}(d) \cite{supMaterial}.
Fitted values for $\tau_i$ are on the order of the $\sim 1$ ps time resolution of our THz probe which limits the determination of exact rise times.
We note that there is no finite fluence threshold for the onset of the $\tau_i$ dynamics since $\Delta \sigma_i$ is nonzero for all fluences. 
Notably, we observe a change $\Delta\sigma_i$ close to $\sim 10^4$ ($\Omega$ cm)$^{-1}$, occurring in $\sim 1$ ps in a 5.7 nm thick film.
Such a conductivity transient is significantly larger and faster than those observed in other IMT materials, such as VO$_2$ and V$_2$O$_3$ \cite{Abreu2017, Hilton2007}.
These fast dynamics correspond to the thermal quench, which sets the stage for the subsequent dynamics shown in Figs. \ref{figOPTP}(b) -  \ref{figOPTP}(d).
As such, to understand the dynamics beyond $\tau_i$ it is important to estimate the final temperature $T_f$, the peak temperature obtained in the sample before it cools down as heat transfers to the substrate, as a function of $T_i$ and $F_{inc}$.

\begin{figure} [htb]
\begin{center}
\includegraphics[width=0.6\textwidth,keepaspectratio=true]{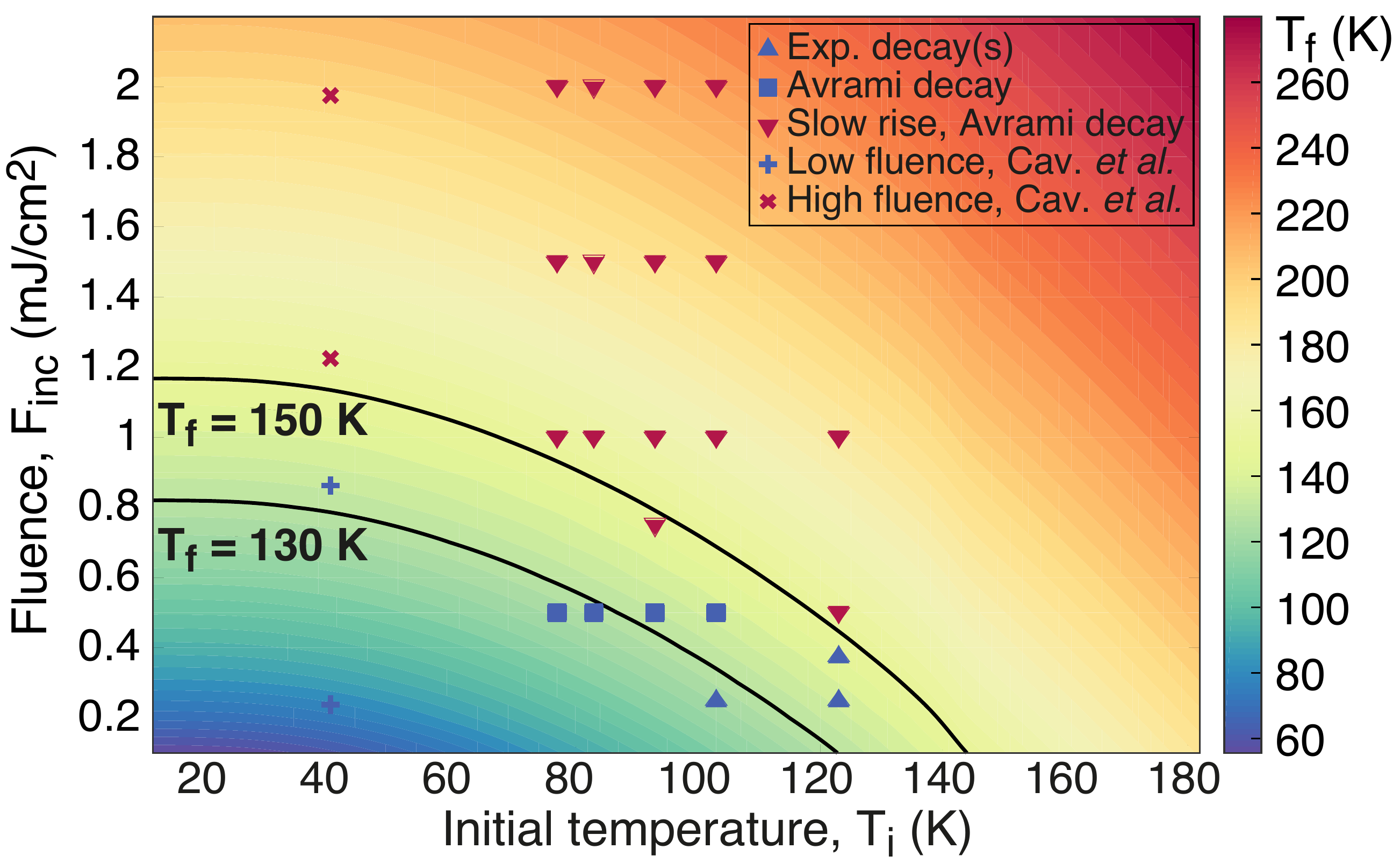}
\caption{
Intensity plot of $T_f$ vs. $T_i$ and $F_{inc}$, in NNO.
The black line at $T_f = 130$ K (150 K) marks the center of the cooling (warming) branch of the IMT, denoting the coexistence region (Fig. \ref{figOPTP}(a)).
Markers distinguish the two regions seen in our $\Delta \sigma(t)$ data (triangles, squares), and in magnetic order dynamics (crosses) by Caviglia \emph{et al.} \cite{Caviglia2013}.
The weak (strong) quench region corresponds to $T_f \simeq T_{IMT}$ ($T_f > T_{IMT}$). 
}
\label{figTTM}
\end{center}
\end{figure}	

$T_f$ can be estimated \cite{supMaterial} from $T_i$, the specific heat \cite{Barbeta2011} and the deposited energy density at a given $F_{inc}$ \cite{Stewart2011}.
Results are shown in Fig. \ref{figTTM} as an intensity map of $T_f$ vs. $T_i$ and $F_{inc}$.
Importantly, calculated $T_f$ values shown in Fig. \ref{figTTM} agree well with the $T_f$ values obtained by comparing $\Delta \sigma_i$ with the static $\sigma (T)$ curve (Fig. \ref{figOPTP}(a)). 
The colored points in Fig. \ref{figTTM} correspond to $T_f$ determined for the data in Figs. \ref{figOPTP}(b) - \ref{figOPTP}(d) (the vertical row of points at $T_i=$ 40 K are estimated from Caviglia \emph{et al.} \cite{Caviglia2013}).
Two regions can be clearly identified, corresponding roughly to the weak thermal quench region ($T_f \simeq T_{IMT}$) and the strong quench region ($T_f > T_{IMT}$).
Red triangles label the strong quench region, where dynamics subsequent to the quench exhibit a conductivity increase followed by non-exponential relaxation (Fig. \ref{figOPTP}(b), $F_{inc} >$ 0.5 mJ/cm$^2$).
The weak quench region for which there is no slow rise of conductivity (Figs. \ref{figOPTP}(c) - \ref{figOPTP}(d)) is labeled with blue symbols, with further refinement to indicate the onset of non-exponential decay (blue squares) versus a simple two-exponent decay (blue pyramids).
The results of Figs. \ref{figOPTP}(b) - \ref{figOPTP}(d) and Fig. \ref{figTTM} reveal that the $\Delta \sigma(t)$ IMT dynamics strongly depend on the magnitude of the quench.

Next, we turn to the $\Delta \sigma(t)$ IMT dynamics of ENO (Fig. \ref{figOPTP}(e)), as an interesting point of comparison to NNO.
The static $\sigma(T)$ (inset of Fig. \ref{figOPTP}(e)) yields $T_{IMT} \simeq 380$ K and exhibits no clear hysteresis.
$\Delta \sigma (t)$ in ENO is measured for $F_{inc} = $ 1 - 3 mJ/cm$^2$ (Fig. \ref{figOPTP}(e) shows $F_{inc}=2$ mJ/cm$^2$) and $T_i = $ 120 - 330 K, both below and above $T_N = 200$ K.
For all $T_i$ values the dynamics show a fast rise and a two-exponent decay (as observed for NNO in the low fluence range, Fig. \ref{figOPTP}(d), where the weak thermal quench is not sufficient to stabilize the metallic phase).
No slow conductivity rise or non-exponential recovery dynamics is observed for any of the $T_i$ and $F_{inc}$ combinations.
This is true even for $T_i=330$ K, at the onset of the static IMT, where photoexcitation places ENO well into the IMT region.
$\Delta \sigma (t)$ in ENO therefore does not exhibit any features suggestive of first-order dynamics, which is consistent with the lack of hysteresis in $\sigma(T)$ (Fig. \ref{figOPTP}(e)) and d.c. transport \cite{Meyers2013} measurements.
The conductivity dynamics are purely thermal with the $\Delta \sigma (t)$ decrease tracking the film temperature as cooling proceeds via heat transfer to the substrate.
The contrast between ENO  and NNO further suggests that the high fluence results for NNO in Fig. \ref{figOPTP}(b) arise from strong first-order dynamics. 

\begin{figure} [htb]
\begin{center}
\includegraphics[width=0.5\textwidth,keepaspectratio=true]{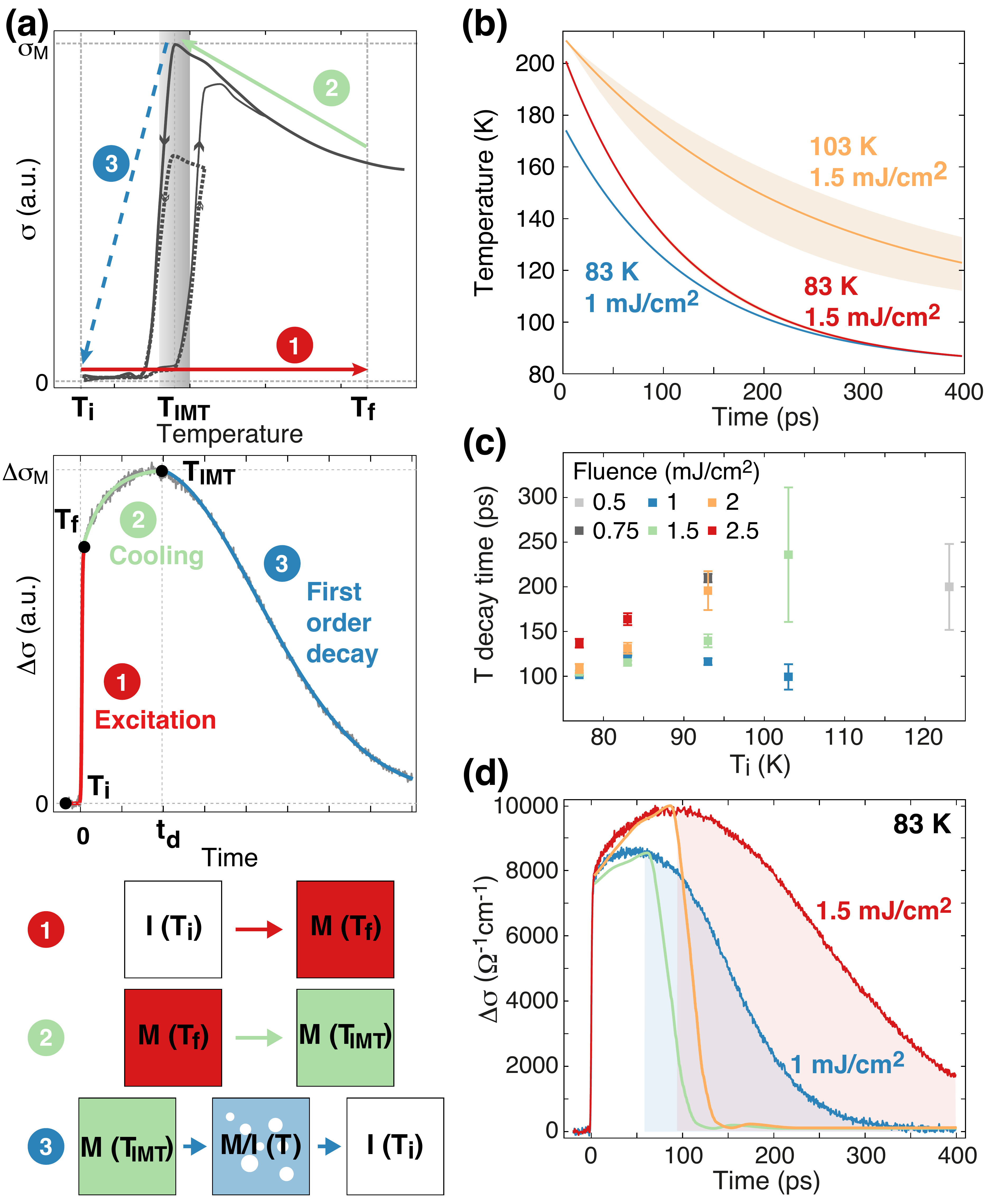}
\caption{
a) Excitation and recovery steps for $T_f > T_{IMT}$:
1) fast quench to  $T_f$;
2) cooling down to $T_{IMT}$, the onset of phase coexistence;
3) first-order non-exponential Avrami recovery.
Top: static mapping of the dynamic data shown in the middle.
Bottom: schematic of the three-step process showing the different sample stages -- insulating at $T_i$ (white), metallic at $T_f$ (red), metallic at $T_{IMT}$ (green) and coexistence at $T < T_{IMT}$ (blue / white for metallic / insulating).
b) Evolution of the film temperature determined from the increase in $\Delta \sigma (t)$  with $\tau_2$ in step 2 (cf. text).
Shaded regions indicate error bars estimated from the fitting error of $t_d$ (smaller than the line thickness for the 83 K traces). 
c) Temperature decay rates, for $T_i$ and $F_{inc}$ values where $t_d$ could be obtained from the fit.
d) Purely thermal recovery expected in the absence of nucleation and growth (orange, green) compared to the actual first-order dynamics (red, blue), at $T_i$ = 83 K and $F_{inc}=1.5$ mJ/cm$^2$ (red, orange) or $F_{inc}=1$ mJ/cm$^2$ (blue, green).
}
\label{figStory}
\end{center}
\end{figure}

We now focus on the NNO post quench dynamics when $T_f$ is well above $T_{IMT}$, corresponding to the data in Fig. \ref{figOPTP}(b) for $F_{inc} >$ 0.5 mJ/cm$^2$.
In Fig. \ref{figStory}(a), we label as step 1 the initial ultrafast increase of $\Delta \sigma (t)$, corresponding to carrier excitation and heating dynamics that quenches the system to $T_f$.
The second increase in $\Delta \sigma (t)$, on a $\tau_2$ timescale, arises from cooling via heat escape from the ultrathin film (step 2) and is consistent with the negative slope of $\sigma(T)$ for $T > T_{IMT}$.
For $T_f \gg T_{IMT}$ the transient conductivity reaches the maximum metallic conductivity ($\sigma_M$, top panel of Fig. \ref{figStory}(a)), such as for the 1.5 mJ/cm$^2$ dynamics plotted in Fig. \ref{figStory}(d) (red).
At lower $T_f $, though still above $T_{IMT}$, $\Delta \sigma (t)$ increases but follows a modified hysteresis loop that never reaches $\sigma_M$ (dashed loop in the top panel of Fig. \ref{figStory}(a) and 1 mJ/cm$^2$ trace in Fig. \ref{figStory}(d) (blue)). 

Importantly, the metallic state conductivity dynamics persist until $t_d$ (see Eq. 1), when $T_{IMT}$ is finally reached.
Up to this point the dynamics are purely thermal.
As such, the sample temperature can be estimated by converting the transient conductivity increase between the fast $\tau_i$ rise and $t_d$ to a transient temperature decrease using the static data of Fig. \ref{figOPTP}(a) \cite{supMaterial}.
In this procedure we use the $\Delta \sigma_i$ and $t_d$ values obtained from the fits to Eq. \ref{eq1}.
The sample temperature decrease dynamics and decay times are shown in Figs. \ref{figStory}(b) and \ref{figStory}(c), with error bars estimated from the fitting errors of $t_d$.
The cooling rate is on the order of 1 K/ps (though it slows down with time given the exponential evolution) and cooling continues unabated below $T_{IMT}$.
Thus, beyond $t_d$ the sample is supercooled into the coexistence region, initiating non-equilibrium first-order phase transition dynamics.

As determined  from the fit of the data in Fig. \ref{figOPTP}(b) to Eq. \ref{eq1}, the conductivity recovery dynamics of NNO are non-exponential.
Indeed, these dynamics are well described by a 2d Avrami model, characterized by a stretched exponential with an exponent of 2, as expected from quickly exhausted nucleation and 2d ballistic growth in thin films \cite{Abreu2015, supMaterial}.
Using the sample temperature determined in Fig. \ref{figStory}(b) we can determine how the conductivity dynamics would proceed for a purely thermal relaxation.
This is plotted as solid lines in Fig. \ref{figStory}(d), revealing a clear discrepancy after $t_d$ (shaded regions).
Crucially, the slow conductivity recovery is due to the growth of the initial insulating phase at the expense of the supercooled metallic regions, which hinders the recovery in comparison to a homogeneous thermal process.


\begin{figure} [htb]
\begin{center}
\includegraphics[width=0.6\textwidth,keepaspectratio=true]{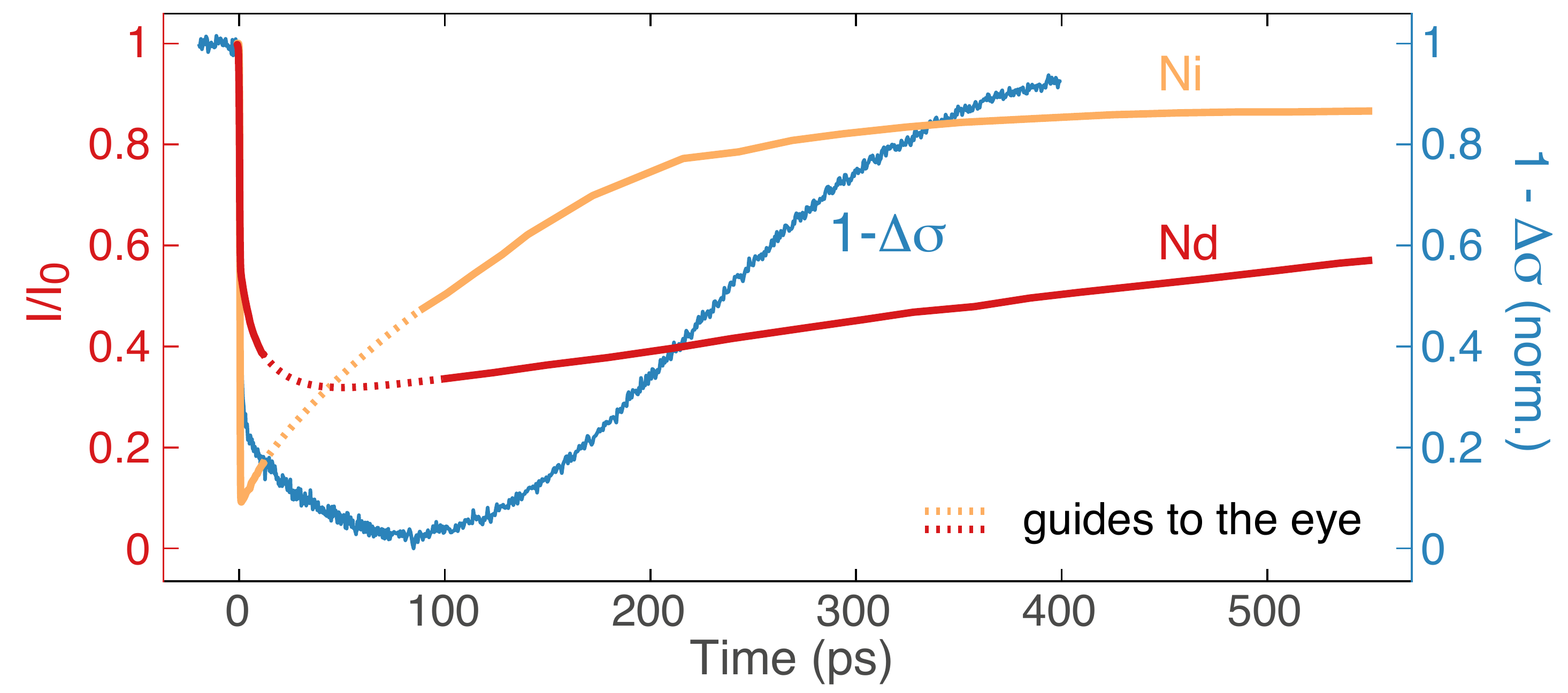}
\caption{
Comparison between conductivity and magnetic dynamics \cite{Caviglia2013}.
The normalized intensity of the magnetic order peak, $I/I_0$, shown in red (orange) for $Nd$ ($Ni$), was measured at $T_i=$ 40 K, $F_{inc}=$ 1.3 mJ/cm$^2$ (1.9 mJ/cm$^2$) \cite{Caviglia2013}.
Dotted sections are guides, not measured data.
The normalized $1-\Delta \sigma$ (blue) was measured at $T_i=$ 77 K, $F_{inc}=$ 1.5 mJ/cm$^2$.
}
\label{figMO}
\end{center}
\end{figure}

Interestingly, $\Delta \sigma (t)$ appears to connect with the ultrafast magnetization dynamics.
Ultrafast resonant soft x-ray diffraction experiments by Caviglia \emph{et al.} \cite{Caviglia2013} have characterized the response of Ni$^{3+}$ and Nd$^{3+}$ magnetic moments, following photoexcitation at 800 nm, for a thicker 100 u.c. NNO film for which $T_{IMT} = 150$ K.
Two regions are reported as a function of fluence (crosses in Fig. \ref{figTTM}), matching the ones we identify here for $\Delta \sigma(t)$.
This is the first indication provided by our comparison of a relationship between electronic and magnetic order during the IMT in NNO.
As a side note, the charge order dynamics reported by Esposito \emph{et al.} \cite{Esposito2018} exhibit only one region corresponding to photoexcitation into the strong thermal quench region ($T_f > T_{IMT}$).

A closer comparison between magnetic and conductivity dynamics is shown in Fig. \ref{figMO} for $T_f > T_{IMT}$.
The recovery of both the magnetic order of the Nd$^{3+}$ ions ($I_{Nd}(t)$) and $1 - \Delta \sigma(t)$ is quite slow, contrary to $I_{Ni}(t)$, the magnetic order of Ni$^{3+}$ ions.
Indeed, it is clear that the slow increase in conductivity ($1 - \Delta \sigma(t)$ decay) is accompanied by a decrease in $I_{Nd}(t)$.
$I_{Ni}(t)$, on the other hand, exhibits a prompt two-exponent recovery, similar to $1 - \Delta \sigma(t)$ for $T_f < T_{IMT}$.
These qualitative conclusions are clear despite differences in the samples and in $T_i$ and $F_{inc}$, and attest to a strong coupling between the electronic and magnetic order of Nd$^{3+}$ in NNO.
Ultrafast magnetic order studies in smaller \emph{R} systems such as ENO can reveal whether first-order signatures are observed in the magnetic dynamics of \emph{R}$^{3+}$ close to $T_N$ or $T_{IMT}$, though absent in $\Delta \sigma(t)$.
Such measurements can complement our conductivity dynamics study and help establish whether the role of $R$ in $R$NiO$_3$ might go beyond the pure steric effect of modifying the perovskite tolerance factor.

In order to characterize the contribution of magnetic and electronic effects during the magnetic and IM transitions in \emph{R}NiO$_3$, theoretical modeling is obviously required.
Previous models were built upon the assumption that charge or bond length disproportionation occur for $T<T_{IMT}$ \cite{Mazin2007, Giovannetti2009, Lau2013, Park2014}, motivated by the charge order and structural transitions observed in the insulating phase of bulk \cite{Alonso2000, Zaghrioui2001} or bulk-like \cite{Staub2002} \emph{R}NiO$_3$ systems.
Such models are able to reproduce one subset of the phases exhibited by bulk \emph{R}NiO$_3$.
Our results support a direction of theoretical investigation that further examines the coupling between magnetic and electronic properties across the IMT \cite{Lee2011}.
It is important to keep in mind that strained ultrathin films differ from bulk; while retaining the characteristic interactions of the system they also pose new experimental and theoretical challenges \cite{Middey2016}.


In summary, we report on the observation of a first-order conductivity recovery in supercooled NNO following photoexcitation above $T_{IMT}$, due to nucleation and growth of the insulating phase which is well described by a 2d Avrami model.
This novel observation is a fundamentally dynamic manifestation of the first order character of the IMT, which the reduced thickness of our films and their fast cooling rate uniquely enable us to detect.
In contrast, the conductivity recovery in ENO follows a purely thermal exponential behavior and exhibits no first-order signatures.
Our analysis emphasizes the critical importance of the coupling between electronic and magnetic degrees of freedom during the IMT in \emph{R}NiO$_3$.
Furthermore, our work highlights the relevance of \emph{R}NiO$_3$ ultrathin films, not only for fundamental studies but also from a technological point of view (NNO exhibits a $10^4$ conductivity increase within $\sim 1$ ps), in ultrafast electronic and magnetic switching applications for which a performance better than that of other TMOs is expected.


E.A., J.Z. and R.D.A. acknowledge support from the DOE - Basic Energy Sciences under Grants DE-FG02-09ER46643 and DE-SC0012375.
E.A. acknowledges support from the Funda\c c\~ao para a Ci\^encia e a Tecnologia, Portugal, through doctoral degree fellowship SFRH/ BD/ 47847/ 2008.
D.M. was supported by the Gordon and Betty Moore Foundation's EPiQS initiative under Grant No. GBMF5307.
X.L. was supported by the DOE under Grant DE-SC0012375.
J.C. was supported by the Gordon and Betty Moore Foundation's EPiQS initiative through Grant No. GBMF4534.


\bibliographystyle{apsrev4-1}
\bibliography{NdNiO3OPTP}



\section*{Supplementary material}

| T | Flu | $\Delta\sigma_i$ | $\tau_i$ | $\Delta\sigma_1$ | $\tau_1$ | $\Delta\sigma_2$ | $\tau_2$ | d | $\tau_{Av}$ | $t_d$ |
|---|---|---|---|---|---|---|---|---|---|---|
| 77 | 0.5 | 4208±22 | 0.36±0.01 | 2676±353 | 45±9 | 0 | 30 | 1±0 | 50±1 | 29±2 |
| | 1 | 6614±38 | 0.46±0.01 | 0 | 30 | 1041±36 | 12±1 | 0.99±0 | 106±1 | 52±1 |
| | 1.5 | 7320±34 | 0.5±0.02 | 0 | 30 | 2287±33 | 31±1 | 0.98±0 | 184±2 | 80±1 |
| | 2 | 7244±32 | 0.54±0.02 | 0 | 30 | 2967±51 | 51±3 | 0.96±0.03 | 301±12 | 99±4 |
| | 2.5 | 6736±28 | 0.55±0.02 | 0 | 30 | 3393±34 | 62±2 | 1 | 438±8 | 145±5 |
| 83 | 0.5 | 4030±41 | 0.35±0.02 | 1752±285 | 30±9 | 0 | 30 | 1 | 59±1 | 28±4 |
| | 1 | 7586±48 | 0.46±0.01 | 0 | 30 | 981±46 | 11±1 | 0.99±0 | 121±1 | 61±1 |
| | 1.5 | 7849±31 | 0.52±0.01 | 0 | 30 | 2168±30 | 33±1 | 0.99±0.01 | 229±3 | 87±2 |
| | 2 | 7407±27 | 0.54±0.02 | 0 | 30 | 2835±34 | 52±2 | 0.94±0.11 | 377±33 | 121±6 |
| | 2.5 | 7093±27 | 0.56±0.02 | 0 | 30 | 3381±29 | 62±2 | 1 | 539±17 | 175±7 |
| 93 | 0.5 | 4775±29 | 0.38±0.01 | 489±50 | 15±4 | 0 | 30 | 0.98±0 | 125±2 | 16±3 |
| | 0.75 | 7953±56 | 0.47±0.02 | 0 | 30 | 284±53 | 11±3 | 0.96±0 | 175±1 | 51±1 |
| | 1 | 8476±41 | 0.51±0.02 | 0 | 30 | 1395±37 | 24±2 | 0.94±0.01 | 262±5 | 76±2 |
| | 1.5 | 7975±35 | 0.53±0.02 | 0 | 30 | 2444±34 | 40±1 | 1 | 554±13 | 128±7 |
| | 2 | 7559±26 | 0.56±0.02 | 0 | 30 | 3165±26 | 67±2 | 1 | 949±126 | 214±23 |
| 103 | 0.25 | 2000±83 | 0.3±0.03 | 1397±83 | 2±0 | -578±31 | 26±2 | 1 | 100000 | 0 |
| | 0.5 | 5306±15 | 0.36±0.01 | 995±354 | 104±32 | 0 | 30 | 0.86±0.01 | 268±6 | 20±12 |
| | 1 | 7974±24 | 0.48±0.01 | 0 | 30 | 1477±26 | 39±2 | 1 | 791±24 | 83±11 |
| | 1.5 | 7748±22 | 0.51±0.01 | 0 | 30 | 2402±22 | 60±1 | 1 | 1686±1016 | 251±79 |
| 123 | 0.25 | 2297±33 | 0.38±0.02 | 599±32 | 6±1 | -933±11 | 131±6 | 1 | 100000 | 0 |

TABLE I. Parameters obtained by fitting the data using Eq. 1, for different temperatures and fluences. The minimum set of timescales was used to fit the data by fixing as many parameters as possible (shown in gray) so that they do not contribute to the fitted function.

## 1. FIT RESULTS

As described in the main text, optical pump - THz probe data were fit using

$$\Delta\sigma_{fit}(t) = \frac{1}{e^{-\frac{t}{\tau_i}}+1}\Big[\Delta\sigma_i + \Delta\sigma_1(e^{-\frac{t}{\tau_1}}-1) - \Delta\sigma_2(e^{-\frac{t}{\tau_2}}-1)\Big]\Big[d\,e^{-\left(\frac{t-t_d}{\tau_{Av}}\right)^2}+(1-d)\Big]. \quad (1)$$

The obtained fitting parameters are shown in Table I. The minimum set of timescales

was used to fit each time trace. In order to use always the same fitting function, some fitting parameters were fixed so that the corresponding timescale does not contribute to the fit. These are shown in gray in Table I.

## 2. CALCULATION OF $T_f$ FROM $T_i$ AND $F_{inc}$

The penetration depth and reflectivity for a photon energy of $E_\gamma = 1.5$ eV are $d = 88$ nm and $R = 0.19$ in NdNiO$_3$, respectively [1, 2], and the insulating gap is $E_g = 0.04$ eV [2]. The energy density deposited by the pump pulse is therefore given by

$$\mathcal{E} = (1 - R) \frac{E_\gamma - E_g}{E_\gamma} \frac{Fluence}{d}.$$

$T_f$ can be estimated from the deposited energy density $\mathcal{E}$ and the specific heat of NdNiO$_3$, $C(T)$ [3]. Since $C(T)$ is temperature dependent, we estimate $T_f$ by calculating the incremental increase in temperature, $\delta T_n = \delta \mathcal{E}/C(T_n)$, as small fractions of $\mathcal{E}$, $\delta \mathcal{E}$, are deposited on the sample at temperature $T_n = T_i + \sum_{m<n} \delta T_m$.

## 3. ESTIMATE OF THE TEMPERATURE DYNAMICS

The temperature dynamics discussed in Figs. 3b and 3c of the main text are calculated assuming a time dependence of the film temperature given by

$$T(t) = (T_f - T_i) \, e^{-\frac{t - t_f}{\tau_T}} + T_i.$$

First, the cooling branch of the static $\sigma(T)$ curve is rescaled so that its maximum equals the maximum value of the dynamic trace, leading to a modified hysteresis loop as shown in the top panel of Fig. 3a. $T_f$ in the equation above is read from the modified $\sigma(T)$ trace as the temperature at which $\Delta \sigma_i$ (from the fit) is obtained. $t_f$ is the time at which $\Delta \sigma_i$ is reached in the $\Delta \sigma(t)$ time trace. Finally, $T_i$ is the initial temperature of the sample, as used in the main text. By construction, $T(t_f) = T_f$. As discussed in the main text, we know that the maximum of $\Delta \sigma(t)$ is obtained at time $t_d$ (extracted from the fit) when the sample has cooled down to $T_{IMT} = 140$ K, so that $T(t_d) = T_{IMT}$. $\tau_T$ can then be obtained, yielding

$$\tau_d = \frac{t_d - t_f}{ln\left(\frac{T_f - T_i}{T_{IMT} - T_i}\right)}.$$

$\tau_d$ values and a few $T(t)$ traces are plotted in Figs. 3c and 3b of the main text, respectively.

Note that Fig. 3d and our conclusions are esssentially unaffected if the thermal model is calculated assuming an exponentially decaying amount of heat instead of an exponentially decaying temperature, taking into account the specific heat and (small) latent heat of the material [3].

## 4. AVRAMI MODEL OF NUCLEATION AND GROWTH

Classical models of nucleation and growth predict a dynamic evolution of the metallic volume fraction $f(t)$, which can be described by the Avrami equation [4]

$$f(t) = 1 - e^{-Kt^n},$$

where K is the rate at which $f(t)$ increases and $n$ is an exponent that depends on the dimensionality and nature of the nucleation and growth process. For quickly exhausted nucleation and 2d ballistic growth one expects $n = 2$ [4]. In this work, a quasi-instantaneous nucleation scenario is quite reasonable given the fast cooling rate which leads to supercooling of the system (cf. Fig. 3 of the main text), and 2d interfacial growth is expected from the thin film geometry of the sample [5]. We therefore use $n = 2$ to describe the dynamical evolution of $f(t)$.

In the phase where both metallic and insulating domains coexist the conductivity of the system ($\sigma$) can be calculated in the Bruggeman effective medium approximation, which relates the conductivity of the metallic domains ($\sigma_m$), that of the insulating domains ($\sigma_i$) and the metallic volume fraction ($f$) [5]. For $\sigma \gg \sigma_i$ a linear relationship arises between $\sigma$ and $f$. This is the case for the transient data we are fitting since the dynamics start from the metallic phase, meaning that $\Delta\sigma(t)$ depends linearly on $f(t)$. It is therefore justified to fit the conductivity recovery dynamics by a stretched exponential of the type $e^{-\left(\frac{t-t_d}{\tau_{Av}}\right)^2}$, as done in Fig. 1 of the main text.



\end{document}